# The effect of undrained heating on a fluid-saturated hardened cement paste

Siavash Ghabezloo[*(1)], Jean Sulem[(1)], Jérémie Saint-Marc[(2)]

(1) Université Paris-Est, UR Navier, CERMES, Ecole Nationale des Ponts et Chaussées, Marne la Vallée, France
(2) TOTAL, Management of Residual Gases Project, Pau, France

## Abstract

The effect of undrained heating on volume change and induced pore pressure increase is an important point to properly understand the behaviour and evaluate the integrity of an oil well cement sheath submitted to rapid temperature changes. This thermal pressurization of the pore fluid is due to the discrepancy between the thermal expansion coefficients of the pore fluid and of the solid matrix. The equations governing the undrained thermo-hydro-mechanical response of a porous material are presented and the effect of undrained heating is studied experimentally for a saturated hardened cement paste. The measured value of the thermal pressurization coefficient is equal to 0.6MPa/°C. The drained and undrained thermal expansion coefficients of the hardened cement paste are also measured in the heating tests. The anomalous thermal behaviour of cement paste pore fluid is back analysed from the results of the undrained heating test.

[*]Corresponding Author: Siavash Ghabezloo, CERMES, Ecole Nationale des Ponts et Chaussées, 6-8 avenue Blaise Pascal, Cité Descartes, 77455 Champs-sur-Marne, Marne la Vallée cedex 2, France
Email: siavash.ghabezloo@cermes.enpc.fr

# 1 Introduction

Temperature increase in saturated porous materials under undrained condition leads to volume change and pore fluid pressure increase. This thermal pressurization of the pore fluid is due to the discrepancy between the thermal expansion coefficients of the pore fluid and of the solid phase. This increase of the pore fluid pressure induces a reduction of the effective mean stress, and can lead to shear failure or hydraulic fracturing. This phenomenon is important in petroleum engineering where the reservoir rock and the well cement lining undergo sudden temperature changes. This is for example the case when extracting heavy oils by steam injection methods where steam is injected into the reservoir to heat the oil to a temperature at which it flows. This rapid increase of temperature could damage cement sheath integrity of wells and lead to loss of zonal isolation. As a consequence, its impact on cement behaviour should be properly understood to be pro-active when designing cement sheath integrity.

The phenomenon of thermal pressurization is also important in environmental engineering for radioactive (exothermal) waste disposal in deep clay geological formations as well as in geophysics in the studies of rapid fault slip events when shear heating tends to increase the pore pressure and to decrease the effective compressive stress and the shearing resistance of the fault material [1].

The values of the thermal pressurization coefficient, defined as the pore pressure increase due to a unit temperature increase in undrained condition, is largely dependent upon the nature of the material, the state of stress, the range of temperature change, the induced damage. The large variability of the thermal pressurization coefficient reported in the literature for different porous materials with values from 0.01MPa/°C to 1.5MPa/°C (see Ghabezloo and Sulem [2] for a review) highlights the necessity of laboratory studies.

The aim of this paper is to study the effect of undrained heating and the phenomenon of thermal pressurization for a fluid-saturated hardened cement paste. Using the framework of thermo-poro-mechanics, the response of hardened cement paste to heating is modelled. An experimental program of drained and undrained heating tests is performed and the tests results are critically discussed.

A better understanding of the effect of undrained heating and induced thermal pressurization phenomenon is an important point to properly understand the behaviour and evaluate the integrity of an oil well cement sheath submitted to rapid temperature changes.

## 2   Thermo-Poro-Mechanical background

We present here the framework used to describe the thermo-elastic volumetric behaviour of a porous material which can be heterogeneous and anisotropic at the micro-scale. The theoretical basis of the formulation has been presented in many earlier studies. Among them, one can refer to the milestone papers and textbooks of Biot and Willis [3], Brown and Korringa [4], Rice and Cleary [5], Palciauskas and Domenico [6], McTigue [7], Zimmerman [8], Detournay and Cheng [9], Vardoulakis and Sulem [10], Coussy [11]. This framework is recalled here in a comprehensive manner in order to clarify the mathematical and physical significance of the different parameters which are measured in our experimental program. We extend here the presentation made in Ghabezloo *et al*. [12] to account for the effect of temperature.

A fluid saturated porous material can be seen as a mixture of two phases: a solid phase and a fluid phase. The solid phase may be itself made up of several constituents. The porosity $\phi$ is defined as the ratio of the volume of the porous space $V_\phi$ to the total volume $V$ in the actual (deformed) state.

$$\phi = \frac{V_\phi}{V} \tag{1}$$

For a saturated material, the volume of the pore space is equal to the volume of the fluid phase. We consider a saturated sample under an isotropic state of stress $\sigma$ (positive in compression). We choose three independent variables for characterizing the volumetric response of a porous material: the pore pressure $p_f$, the differential pressure $\sigma_d$ as defined below and the temperature $T$.

$$\sigma_d = \sigma - p_f \tag{2}$$

The expression of the variations of the total volume $V$ and of the pore volume $V_\phi$ introduces six parameters:

$$\frac{dV}{V} = \frac{1}{V}\left(\frac{\partial V}{\partial \sigma_d}\right)_{p_f,T} d\sigma_d + \frac{1}{V}\left(\frac{\partial V}{\partial p_f}\right)_{\sigma_d,T} dp_f + \frac{1}{V}\left(\frac{\partial V}{\partial T}\right)_{p_f,\sigma_d} dT \quad (3)$$

$$\frac{dV_\phi}{V_\phi} = \frac{1}{V_\phi}\left(\frac{\partial V_\phi}{\partial \sigma_d}\right)_{p_f,T} d\sigma_d + \frac{1}{V_\phi}\left(\frac{\partial V_\phi}{\partial p_f}\right)_{\sigma_d,T} dp_f + \frac{1}{V_\phi}\left(\frac{\partial V_\phi}{\partial T}\right)_{p_f,\sigma_d} dT$$

$$\frac{1}{K_d} = -\frac{1}{V}\left(\frac{\partial V}{\partial \sigma_d}\right)_{p_f,T} \quad , \quad \frac{1}{K_p} = -\frac{1}{V_\phi}\left(\frac{\partial V_\phi}{\partial \sigma_d}\right)_{p_f,T} \quad (4)$$

$$\frac{1}{K_s} = -\frac{1}{V}\left(\frac{\partial V}{\partial p_f}\right)_{\sigma_d,T} \quad , \quad \frac{1}{K_\phi} = -\frac{1}{V_\phi}\left(\frac{\partial V_\phi}{\partial p_f}\right)_{\sigma_d,T} \quad (5)$$

$$\alpha_d = \frac{1}{V}\left(\frac{\partial V}{\partial T}\right)_{p_f,\sigma_d} \quad , \quad \alpha_\phi = \frac{1}{V_\phi}\left(\frac{\partial V_\phi}{\partial T}\right)_{p_f,\sigma_d} \quad (6)$$

Equation (4) corresponds to an isotropic and isothermal drained compression test in which the pore pressure is controlled to remain constant in the sample. The variations of the total volume of the sample $V$ and of the volume of the pore space $V_\phi$ with respect to the applied confining pressure give the (isothermal) drained bulk modulus $K_d$ and the modulus $K_p$. Equation (5) corresponds to the so-called unjacketed isothermal compression test, in which equal increments of confining pressure and pore pressure are simultaneously applied to the sample, as if the sample is submerged, without a jacket, into a fluid under the pressure $p_f$. The differential pressure $\sigma_d$ in this condition remains constant. Neglecting the deformation of the jacket, the measured volumetric strain with the applied pressure gives the unjacketed modulus $K_s$. The variation of the pore volume of the sample in this test, evaluated from the quantity of fluid exchanged between the sample and the pore pressure generator when applying equal increments of confining pressure and pore pressure could in principle give the modulus $K_\phi$. However experimental evaluation of this parameter is very difficult as the volume of the exchanged fluid has to be corrected for the effect of fluid compressibility and also for the effect of the deformations of the pore pressure generator and drainage system in order to access to the variation of the pore volume of the sample. In the case of a porous material which is homogeneous and isotropic at the micro-scale, the sample would deform in an unjacketed test as if all the pores were filled with the solid component. The skeleton and the solid component experience a uniform volumetric strain with no change of the porosity [9]. For such a material $K_s = K_\phi = K_m$, where $K_m$ is the bulk modulus of the single solid constituent of the porous material. In the case of a porous material which is composed of two or more solids and therefore is heterogeneous at the micro-scale, the unjacketed modulus $K_s$ is some weighted average of the bulk moduli of solid constituents [13]. The modulus $K_\phi$ for such a material has a complicated dependence on material properties. Generally it is not bounded by the elastic moduli of the solid components and can even have a negative sign if the bulk moduli of the individual solid components are greatly different one from another [14][15].

Equation (6) corresponds to a drained heating test in which the pore pressure is controlled to remain constant in the sample while the thermal loading is applied. The variations of the total volume of the sample $V$ and of the volume of the pore space $V_\phi$ with respect to the applied thermal loading give the volumetric drained thermal expansion coefficient $\alpha_d$ and pore volume thermal expansion coefficient $\alpha_\phi$. According to Palciauskas and Domenico [6], the coefficient $\alpha_d$ measured in a heating test also takes into account the non-reversible thermal deformations which can be produced by the microfracture generation due to the discrepancy between the thermal expansions of different constituents of the porous material, but the experimental study of Walsh [16] on rocks showed that the microfracture generation is only initiated at elevated temperatures. Like for $K_\phi$, the experimental evaluation of $\alpha_\phi$ is very difficult as the volume of the exchanged fluid has to be corrected for the effect of thermal expansion of the fluid and also for the effect of the thermal deformations of the pore pressure generator and of the drainage system in order to access to the variation of the pore volume of the sample. In the case of a micro-homogeneous and micro-isotropic porous material, $\alpha_d = \alpha_\phi = \alpha_m$, where $\alpha_m$ is the thermal expansion coefficient of the single solid constituent of the porous material. For such a material, there is no change of porosity during a drained thermal loading because an isotropic thermal expansion would cause a proportional change in every linear dimension of the body. In the general case, the difference between the expansion coefficients $\alpha_d$ and $\alpha_m$ reflects the difference between the thermal response of the bulk porous medium and that of the solid phase alone [7].

Using Betti's reciprocal theorem the following relation between the elastic moduli is obtained [4]:

$$\frac{1}{K_p} = \frac{1}{\phi}\left(\frac{1}{K_d} - \frac{1}{K_s}\right) \qquad (7)$$

Using equation (7), the number of the required parameters to characterize the volumetric thermo-elastic behaviour of a porous material is reduced to five.

The variations of the total volume $V$ and of the pore volume $V_\phi$ (equation (3)) can be rewritten as follows:

$$\frac{dV}{V} = -\frac{d\sigma_d}{K_d} - \frac{dp_f}{K_s} + \alpha_d dT$$

$$\frac{dV_\phi}{V_\phi} = -\frac{d\sigma_d}{K_p} - \frac{dp_f}{K_\phi} + \alpha_\phi dT \qquad (8)$$

The incremental volumetric strain $d\varepsilon = -dV/V$ is thus expressed as:

$$d\varepsilon = -\frac{dV}{V} = \frac{d\sigma_d}{K_d} + \frac{dp_f}{K_s} - \alpha_d dT \qquad (9)$$

Using the definition of the porosity presented in equation (1), the following equation is obtained for the variation of the porosity:

$$\frac{d\phi}{\phi} = \frac{dV_\phi}{V_\phi} - \frac{dV}{V} \qquad (10)$$

Replacing equations (8) and (7) in equation (10), the expression of the variation of porosity is obtained:

$$\frac{d\phi}{\phi} = -\frac{1}{\phi}\left(\frac{1-\phi}{K_d} - \frac{1}{K_s}\right)d\sigma_d + \left(\frac{1}{K_s} - \frac{1}{K_\phi}\right)dp_f - \left(\alpha_d - \alpha_\phi\right)dT \qquad (11)$$

Equation (11) clearly shows that in the case of an ideal porous material ($\alpha_d = \alpha_\phi$, $K_s = K_\phi$) there would be no change of porosity during an isotropic thermal loading or during an unjacketed loading ($d\sigma_d = 0$).

The undrained condition is defined as a condition in which the mass of the fluid phase is constant ($dm_f = 0$). Under this condition we choose three different independent variables: the total stress $\sigma$, the fluid mass $m_f$, and the temperature $T$. The measured quantities are the total volume $V$ and the pore pressure $p_f$. Writing the expression of the variation of these quantities, we can define four new parameters to describe the response of the porous material in undrained condition:

$$B = \left(\frac{\partial p_f}{\partial \sigma}\right)_{m_f,T} \quad , \quad \frac{1}{K_u} = -\frac{1}{V}\left(\frac{\partial V}{\partial \sigma}\right)_{m_f,T} \tag{12}$$

$$\Lambda = \left(\frac{\partial p_f}{\partial T}\right)_{m_f,\sigma} \quad , \quad \alpha_u = \frac{1}{V}\left(\frac{\partial V}{\partial T}\right)_{m_f,\sigma} \tag{13}$$

The parameter $K_u$ is the undrained bulk modulus and $B$ is the so-called Skempton coefficient [17]. $\alpha_u$ is the undrained volumetric thermal expansion coefficient and $\Lambda$ is the thermal pressurization coefficient. Fluid mass conservation under undrained condition leads to the following expression for variation of the volume of the fluid:

$$-\frac{dV_\phi}{V_\phi} = \frac{dp_f}{K_f} - \alpha_f dT \tag{14}$$

where $K_f$ and $\alpha_f$ are respectively the fluid compression modulus and thermal expansion coefficient. Replacing equation (14) in equation (8) and using equations (2) and (7) the following expression is found for the variation of pore pressure with the confining pressure and temperature in undrained condition:

$$dp_f = \frac{(1/K_d - 1/K_s)}{(1/K_d - 1/K_s) + \phi(1/K_f - 1/K_\phi)} d\sigma + \frac{\phi(\alpha_f - \alpha_\phi)}{(1/K_d - 1/K_s) + \phi(1/K_f - 1/K_\phi)} dT \tag{15}$$

Comparing equation (15) with the definitions of the Skempton coefficient $B$ and of the thermal pressurization coefficient $\Lambda$ (equations (12) and (13)), we obtain:

$$B = \frac{(1/K_d - 1/K_s)}{(1/K_d - 1/K_s) + \phi(1/K_f - 1/K_\phi)} \tag{16}$$

$$\Lambda = \frac{\phi(\alpha_f - \alpha_\phi)}{(1/K_d - 1/K_s) + \phi(1/K_f - 1/K_\phi)} \tag{17}$$

Equation (17) clearly highlights that the thermal pressurization of porous materials is caused by the discrepancy between the thermal expansion of the pore fluid and the one of the pore volume.

The variation of the total volume in undrained condition is given by the undrained bulk modulus $K_u$ and the undrained thermal expansion coefficient $\alpha_u$ as presented in equations (12) and (13). Replacing $dV/V = -d\sigma/K_u + \alpha_u dT$, $d\sigma_d = d\sigma - dp_f$ and $dp_f = Bd\sigma + \Lambda dT$ in equation (9), the following relationships are found between the various coefficients:

$$K_u = \frac{K_d}{1 - B(1 - K_d/K_s)} \qquad (18)$$

$$\alpha_u = \alpha_d + \Lambda(1/K_d - 1/K_s) \qquad (19)$$

Using equations (16) and (17) in equation (19) the following expression as presented by McTigue [7] for the undrained thermal expansion coefficient is retrieved:

$$\alpha_u = \alpha_d + \phi B(\alpha_f - \alpha_\phi) \qquad (20)$$

## 2.1 Influence of non-elastic strains

The above framework can be extended to account for the effect of non-elastic strains. These strains can be plastic, viscoelastic or viscoplastic and induce non-elastic porosity changes. The non-elastic changes of the total volume, pore volume and solid volume are defined by:

$$dV^{ne} = dV - dV^e \; ; \; dV_\phi^{ne} = dV_\phi - dV_\phi^e \; ; \; dV_s^{ne} = dV_s - dV_s^e \qquad (21)$$

The non-elastic increment of pore volume $dV_\phi^{ne}$ can be calculated from the definition of the porosity (equation (1)) and knowing that $V_\phi = V - V_s$.

$$dV_\phi^{ne} = dV_\phi - dV_\phi^e = V\left[-d\varepsilon^{ne} + (1-\phi)d\varepsilon_s^{ne}\right] \qquad (22)$$

From (22) we obtain:

$$\frac{dV_\phi^{ne}}{V_\phi} = \frac{-1}{\phi}\left[d\varepsilon^{ne} - (1-\phi)d\varepsilon_s^{ne}\right] \qquad (23)$$

Using equation (23), equation (8) is re-written with the additional contribution of the non-elastic volume changes:

$$-\frac{dV}{V} = \frac{d\sigma_d}{K_d} + \frac{dp_f}{K_s} - \alpha_d dT + d\varepsilon^{ne}$$

$$-\frac{dV_\phi}{V_\phi} = \frac{d\sigma_d}{K_p} + \frac{dp_f}{K_\phi} + \frac{d\varepsilon^{ne}}{\phi} - \alpha_\phi dT + \frac{d\varepsilon^{ne}}{\phi} - \frac{1-\phi}{\phi} d\varepsilon_s^{ne}$$

(24)

Using equations (10) and (24) the following relation is obtained for the variations of the porosity:

$$\frac{d\phi}{\phi} = -\frac{1}{\phi}\left(\frac{1-\phi}{K_d} - \frac{1}{K_s}\right)d\sigma_d + \left(\frac{1}{K_s} - \frac{1}{K_\phi}\right)dp_f - (\alpha_d - \alpha_\phi)dT - \frac{1-\phi}{\phi}\left(d\varepsilon^{ne} - d\varepsilon_s^{ne}\right) \quad (25)$$

Replacing equation (14) in equation (24) and using equations (2) and (7), the following expression is found for the variation of pore pressure with the confining pressure and temperature in undrained condition, in presence of non-elastic volume changes:

$$dp_f = Bd\sigma + \Lambda dT + \frac{d\varepsilon^{ne} - (1-\phi)d\varepsilon_s^{ne}}{(1/K_d - 1/K_s) + \phi(1/K_f - 1/K_\phi)} \quad (26)$$

Equation (26) shows that non-elastic volume changes add an additional term in the generated pore pressure. In the case of a material for which the solid phase is elastic ($d\varepsilon_s^{ne} = 0$), equation (26) can be rewritten as:

$$dp_f = Bd\sigma + \Lambda dT + \frac{d\varepsilon^{ne}}{(1/K_d - 1/K_s) + \phi(1/K_f - 1/K_\phi)} \quad (27)$$

## 2.2 Cement paste porosity to be used for poromechanical calculations

The size of pores in the microstructure of hardened cement paste covers an impressive range, from nanometre-sized gel pores, to micro-metre sized capillary pores and millimetre-sized air voids [18]. From the different microstructural models of C-S-H, it can be seen that a part of the water in the pore structure of cement paste is interlayer structural water. Feldman and Sereda [19] propose a model for multilayer structure of C-S-H that postulates the existence of interlayer space containing strongly adsorbed water. According to Feldman [20], the interlayer water behaves as a solid bridge between the layers and consequently the interlayer space can not be included in the porosity and the interlayer water must be regarded as a part of the solid structure of hydrated cement paste. Feldman's experiments show that the interlayer water evaporates at very low relative humidity, below 11%. This can be seen also in Jennings' [21][22] microstructural model of C-S-H, in which the amorphous colloidal structure of the C-S-H is organized in elements, called 'globules'. In Jennings' [22] model, the globule is composed of solid C-S-H sheets, intra-globule porosity and a monolayer of water on the surface. For relative humidities below 11%, a part of the water filling the intra-globule porosity is evaporated. In a porosity measurement procedure in which the sample is dried at 105°C until a constant mass is achieved, a part of the interlayer water is evaporated and is thus included in the measured porosity [18][23]. Consequently, the porosity measured by oven-drying at 105°C should not be used in poromechanical formulation, as it includes a part of the interlayer porosity. The free-water porosity is defined and measured by equilibrating the cement paste at 11% relative humidity [18][23]. Some measurements of the free-water porosity, which is obviously smaller than the total porosity measured by drying at 105°C, from Feldman [24] are presented by Jennings *et al*. [18]. From the above discussion it appears that the free-water porosity is the cement paste porosity that should be used in the poromechanical formulations. This is in accordance with the assumption made by Ulm *et al*. [25] in their multi-scale microstructural model for the evaluation of the poromechanical properties of hardened cement paste. These authors use the Jennings' [21] model and exclude the intra-globule porosity from the cement paste porosity used in the calculations. According to Ulm *et al*. [25], considering the characteristic size of the interlayer space, which is less than ten water molecules in size, it is recognized that the water in this space can not be regarded as a bulk water phase. On the other hand, Scherer *et al*. [26] consider that the pores with a diameter less than 2nm, which corresponds to the size of 8 water molecules, can participate in water transfer phenomenon for the permeabilities in the range of nanodarcy (10-21m2). According to these authors, on the surface of the solid, the thickness of the layer of immobile water is about 0.5nm which corresponds to two water molecules. Accordingly, these authors reduce the total porosity of the cement paste for the effect of the thickness of the immobile water (see equation 17 in Scherer *et al*. [26] ).

In Ghabezloo *et al*. [12], the analysis of the experimentally evaluated poroelastic parameters of the hardened cement paste clearly shows that the porosity that should be used in poromechanical calculations is smaller than the total porosity. This statement is based on the following inequality [13]:

$$\frac{1}{K_\phi} \leq \frac{1 - K_d/K_s}{\phi K_s} \tag{28}$$

Replacing this inequality in the expression of $1/K_\phi$ obtained from equation (16) the following upper limit is obtained for the cement paste porosity:

$$\phi \leq K_f \left( \frac{1}{K_d} - \frac{1}{K_s} \right) \left( \frac{K_d}{K_s} + \frac{1-B}{B} \right) \quad (29)$$

Replacing the values of poroelastic parameters, $K_d = 8.7\text{GPa}$, $K_s = 21\text{GPa}$ and $B = 0.4$ evaluated in experimental study of Ghabezloo *et al.* [12], and taking $K_f = 2.2\text{GPa}$, from equation (29) we obtain $\phi \leq 0.28$ which is smaller than the total porosity of the tested cement paste, measured by oven drying at 105°C equal to 0.35. This analysis is in accordance with the discussion presented above and confirms that the porosity of the cement paste that should be used in poromechanical calculations is smaller than the total porosity.

Taylor [23] (in his Fig. 8.5) presents the values of total and free-water porosity derived from calculated phase compositions of mature ordinary cement paste with varying w/c ratio. The comparison of the calculated free-water porosity with the porosity measured by mercury intrusion shows that for w/c ratios smaller than 0.5, the free-water porosity can be approximated by the mercury porosity. For higher w/c ratios, the free-water porosity is somewhat higher than the mercury porosity. This approximation of the free-water porosity with the mercury porosity is made in Ghabezloo *et al.* [12] and a very good compatibility of the experimental results is obtained.

## 3   Experimental program

In order to evaluate the effect of temperature on the behaviour of hardened cement paste a drained and an undrained heating test are performed and presented in the following.

### 3.1  *Sample preparation*

A class G oil well cement was used to prepare the cement paste with a water to cement ratio $w/c = 0.44$. Two additives, a dispersant and an anti-foaming agent were used in the paste. The fresh paste is conserved in 14cm cubic moulds for four days in lime saturated water at 90°C temperature. After this period the temperature is reduced slightly to prevent the cracking of the blocs due to a sudden temperature change. Then, the blocs were cored and cut to obtain cylindrical samples with 38mm diameter and 76mm length. The two ends of the cylindrical samples were rectified to obtain horizontal surfaces. After the preparation of the samples, the geometry and the weight of the samples were measured. To insure the homogeneity and the integrity of the samples, measurements of wave velocity and dynamic elastic modulus were performed on all of them. These measurements were performed in ambient temperature.

After the sample preparation phase, the samples have been submerged in a fluid which is neutral regarding to the pore fluid of the cement paste to prevent chemical reactions during the period of curing. The composition of the natural fluid, presented in Table (1), is provided by TOTAL and is commonly used in the laboratory experiments of the company. The samples were cured for at least three months at 90°C in the neutral fluid with pH=13 before performing the tests. Before doing each test, the temperature of the sample was reduced slowly to prevent any thermal cracking.

| Material | Quantity (g) in 1kg pure water |
|---|---|
| $Al(OH)_3$ | 7.8 |
| $Ca(OH)_2$ | 7.4 |
| NaCl | 0.3 |
| NaOH | 5.93 |
| $Na_2SO_4$ | 7.6 |
| $SiO_2$ | 1.2 |
| pH ≈ | 13 |

**Table 1- Composition of neutral fluid**

The porosity of the samples is studied by two methods: oven drying and mercury intrusion porosimetry. The total porosity is measured by drying the samples at 105°C until a constant mass is achieved, and an average value equal to $\phi = 0.35$ is obtained. As mentioned in section 2.2, this porosity includes a part of the interlayer space of the cement paste. The mercury intrusion porosimetry is performed on the samples which are dried before the tests with the freeze-drying technique using liquid nitrogen which is, according to Gallé [27], the most suitable drying procedure to investigate the pore structure of cement-based materials. With a maximum intruding pressure of 200MPa the average mercury porosity of the samples is obtained equal to $\phi = 0.26$ (Figure (1)). Using the Washburn-Laplace equation ($P = -4\gamma \cos\theta / d$) and assuming a contact angle $\theta$ of 130° and a surface tension of mercury $\gamma$ of 0.483N/m from Ref.[23], the maximum intruding pressure $P$ of 200MPa corresponds to a minimum pore diameter of about 6nm. Based on the discussions of section 2.2, this value of $\phi = 0.26$ will be used as an approximation of the free-water porosity of the studied cement paste in the following poromechanical calculations, as we did also in the study of poromechanical properties of the studied hardened cement paste, presented in Ghabezloo *et al*. [12].

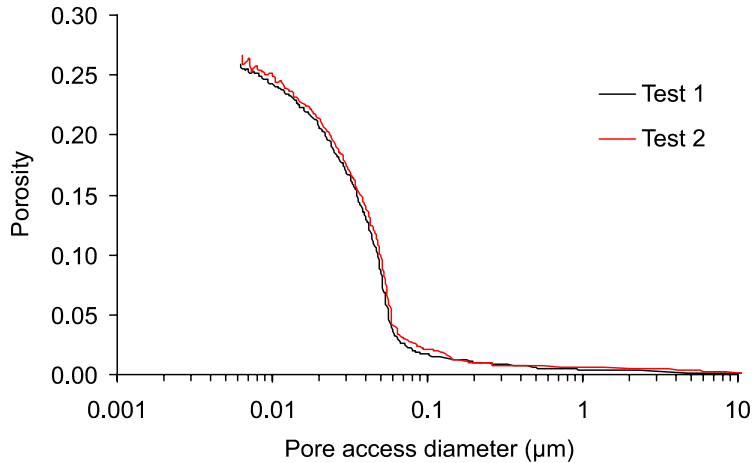

**Figure 1- Mercury intrusion porosimetry tests**

## *3.2 Experimental setting*

The triaxial cell used in this study can sustain a confining pressure up to 60MPa. It contains a system of hydraulic self-compensated piston. The loading piston is then equilibrated during the confining pressure build up and directly applies the deviatoric stress. The axial and radial strains are measured directly on the sample inside the cell with two axial transducers and four radial ones of LVDT type. The confining pressure is applied by a servo controlled high pressure generator. Hydraulic oil is used as confining fluid. The pore pressure is applied by another servo-controlled pressure generator with possible control of pore volume or pore pressure.

The heating system consists of a heating belt around the cell which can apply a temperature change with a given rate and regulate the temperature, and a thermocouple which measures the temperature of the heater. In order to limit the temperature loss, an insulation layer is inserted between the heater element and the external wall of the cell. A second insulation element is also installed beneath the cell. The heating system heats the confining oil and the sample is heated consequently. Therefore there is a discrepancy between the temperature of the heating element in the exterior part of the cell and that of the sample. In order to control the temperature in the centre of the cell, a second thermocouple is placed at the vicinity of sample. The temperature given by this transducer is considered as the sample temperature in the analysis of the test results. Further details and a schematic view of this triaxial cell are presented in Ghabezloo and Sulem [2] and Sulem and Ouffroukh [28].

### 3.2.1 Effect of mechanical and thermal deformation of the drainage system

The undrained condition is defined as a condition in which there is no change in the fluid mass of the porous material. Achieving this condition in a conventional triaxial system is very difficult. In these systems an undrained test is usually performed by closing the valves of the drainage system. As the drainage system is compressible, it experiences volume changes with pressure and temperature changes. A variation of the volume of the drainage system induces a fluid flow into or out of the sample to achieve pressure equilibrium between the sample and the drainage system. This fluid mass exchange between the sample and the drainage system and more generally the mechanical and thermal deformations of the drainage system modify the measured pore pressure during the test.

Ghabezloo and Sulem [2] have presented an extension of the formulation of Bishop [29] and have proposed a simple method for the correction of pore pressure measurements during an undrained heating test. This method requires only two simple calibration tests to determine the needed parameters. The resulting equation which links the corrected value $\Lambda_{cor}$ of the thermal pressurization coefficient to the measured value $\Lambda_{mes}$ can be written as:

$$\Lambda_{cor} = \frac{\Lambda_{mes}}{1 + \frac{\beta(V_L \alpha_{fL} - \alpha_L)}{\phi V (\alpha_f - \alpha_\phi)} - \Lambda_{mes} \frac{V_L c_{fL} + c_L}{\phi V (\alpha_f - \alpha_\phi)}} \tag{30}$$

In equation (30), $V_L$ is the volume of fluid in the drainage system, $V$ is the total volume of the sample, $c_L$ and $\alpha_L$ are the compressibility and the thermal expansion coefficient of the drainage system respectively defined as the variation of the volume of drainage system due to a unit variation in pore pressure and temperature. $\beta = \Delta T_L / \Delta T$ where $\Delta T$ is the temperature change of the sample and $\Delta T_L$ is the equivalent temperature change of the drainage system. $\alpha_{fL}$ and $c_{fL}$ are the thermal expansion coefficient and the compressibility of fluid in the drainage system. As the physical properties of water are temperature and pressure dependent, the effect of temperature and pressure changes on these coefficients has to be taken into account in the analysis of the tests results. Detailed evaluation of these parameters in the calibration tests are presented in Ghabezloo and Sulem [2]. The evaluated parameters for the triaxial system used in this study are:

$$V_L = 2300\,\text{mm}^3$$
$$c_L = 0.27\,\text{mm}^3/\text{MPa}$$
$$\alpha_L = 0.31\,\text{mm}^3/°\text{C}$$
$$\beta = 0.46$$

(31)

## 3.3 Drained heating test

In order to measure the drained thermal expansion coefficient $\alpha_d$ of the hardened cement paste, a drained heating test was carried out under a constant confining pressure of 1.5MPa and a constant pore fluid pressure of 1.0MPa. During the test, the temperature was increased from 18°C to 87°C at a rate of 0.08°C/min and the drained thermal expansion coefficient is evaluated as the slope of the temperature-volumetric strain response (Figure 2).

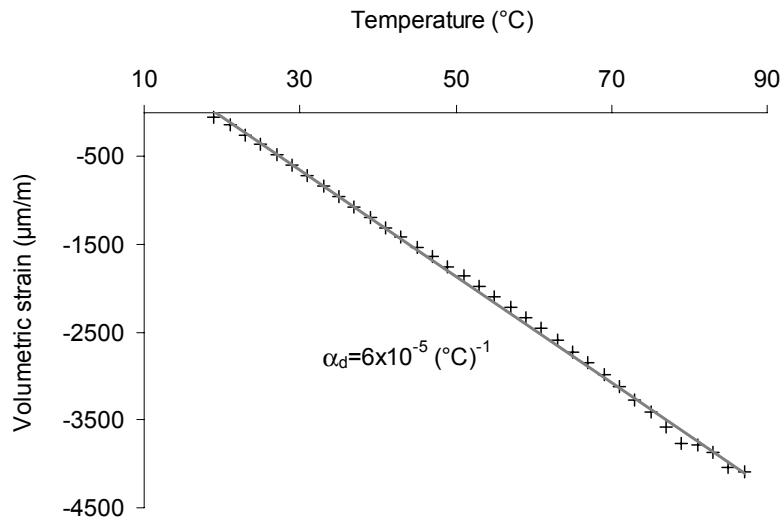

**Figure 2- Drained heating test, temperature - volumetric strain response**

A kinetics analysis of the used heating rate, presented in Appendix 1, shows that with the applied heating rate, the homogeneity of temperature in the sample is quasi instantaneous and the homogeneity of pore pressure along the height of the sample is satisfactory to guaranty that the performed heating test is indeed in drained condition. For drained heating tests performed in the triaxial cell, the test is done using a constant imposed pore pressure and a confining pressure which is greater than the pore pressure. When using a slow rate of temperature change, the duration of the test is large (more than 14 hours in the performed test) and creep may influence the test results. The effect of creep is reduced by choosing small values of imposed confining pressure and pore pressure and a small difference, equal to 0.5MPa, between them. The drained thermal expansion coefficient of hardened cement paste in the performed test is found equal to $6\times10^{-5}\left(°C\right)^{-1}$ (Figure 2).

## *3.4 Undrained heating test*

The phenomenon of thermal pressurization was studied in an undrained heating test under a constant isotropic stress equal to 19MPa. After the saturation phase, the confining pressure is increased up to 19MPa in drained condition at a rate of 0.025MPa/min. In Ghabezloo *et al.* [12] the effect of creep of hardened cement paste under isotropic loading was mentioned. According to equation (26), creep deformations can influence the measured pore pressure variation during the undrained heating test. In order to minimize these effects, the sample was kept under constant load at ambient temperature during about two days before starting the undrained heating test. The volumetric response of the sample during this period is presented in Figure (3) and one can see that the creep strains are stabilized. However, it is well-known that the cement paste creep increases with temperature [30], consequently some additional creep strains may occur during the test due to the temperature increase.

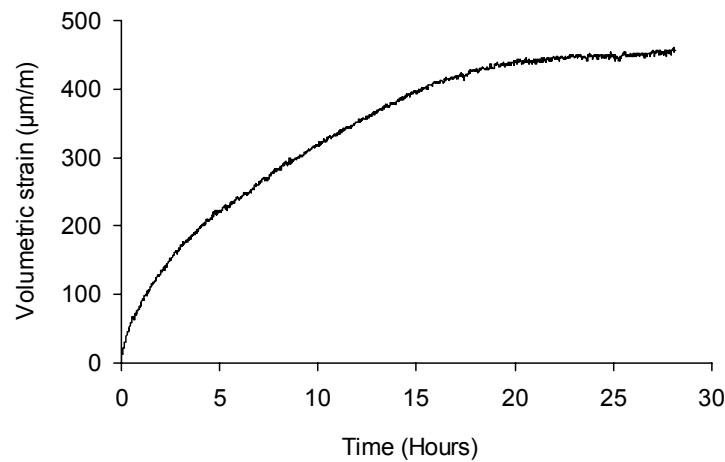

**Figure 3- Creep phase of the tested sample before the undrained heating test**

After the stabilization of the creep strains the temperature was increased at a rate equal to 0.1°C/min and the pore pressure change was monitored during the test. As it can be seen in Appendix 1, with the applied heating rate, the homogeneity of the temperature and consequently the homogeneity of the generated pore pressure in the sample is quasi instantaneous. As in a triaxial device the pore pressure cannot exceed the confining pressure, the heating phase was stopped when the pore pressure reached the confining pressure and the temperature was then decreased. The test results are shown on Figures (4) and (5) where the measured pore pressure variation and the volumetric strain are plotted versus the temperature change.

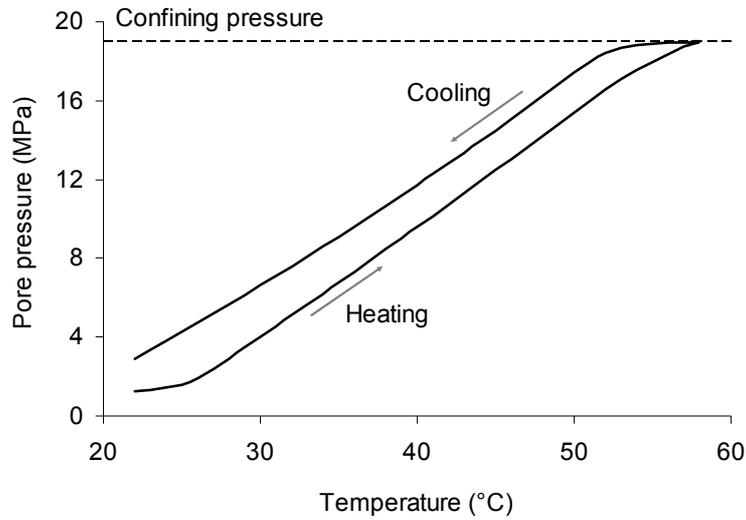

**Figure 4- Undrained heating test, temperature-pore pressure response**

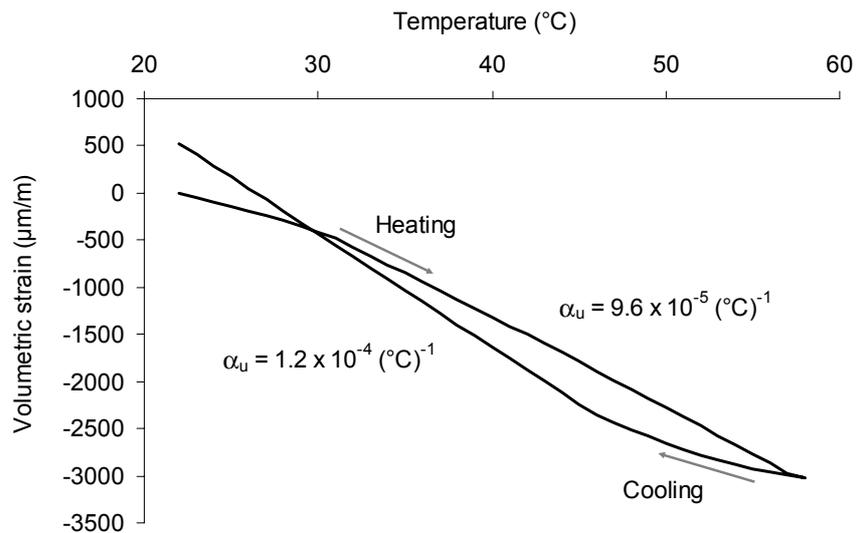

**Figure 5- Undrained heating test, temperature-volumetric strain response**

As can be seen in Figure (4), for a pore pressure close to the confining pressure, the pressurization curve becomes almost horizontal and no more fluid pressurization is observed. This phenomenon is due to the presence of fluid between the sample and the rubber membrane when the difference between the confining pressure and the pore pressure is too small. For the same reason, the pore pressure reduction is delayed at the beginning of the cooling phase.

The measured volumetric strains, presented in Figure (5), show the expansion and the contraction of the sample during the heating and cooling phases respectively.

We can see in Figure (4) that the measured pore pressure at the end of the cooling phase is larger than the one measured at the beginning of the test. The difference between these two values is equal to 1.7MPa. As it can be seen in Figure (5), at the end of the cooling phase a small positive irreversible volumetric strain, equal to 530µm/m is observed. This irreversible volumetric strain can be attributed to creep and may be responsible for the difference between the pore pressures at the beginning and at the end of the undrained heating test. This can be examined quantitatively by introducing the effect of this irreversible volumetric strain (530µm/m) on the pore pressure in equation (27). Based on the results of Ghabezloo *et al.* [12], the following values are considered: $K_d = 8.7\,\text{GPa}$, $K_s = 21\,\text{GPa}$ and $K_\phi = 16.9\,\text{GPa}$. At 40°C which is the average temperature at which the undrained heating test is performed, the compression modulus of the pore fluid can be taken equal to the one of pure water: $K_f = 2.27\,\text{GPa}$ [31]. Inserting these values in equation (27) and with $\phi = 0.26$, the value of pore pressure increase due to the measured irreversible strain is found equal to 3.2MPa which is a bit higher than the measured pore pressure difference between the beginning and the end of the heating cycle, equal to 1.7MPa. This difference can be explained by the leakage of a small quantity of the pore fluid at the end of the heating test when the pore pressure approaches the confining pressure as mentioned above. From simple poroelastic calculations, one can evaluate the quantity of the leaked fluid sufficient to obtain this pore pressure difference, equal to about 17mm3, which is a very small quantity.

### 3.4.1 Heating phase

The undrained thermal expansion coefficient for the heating phase is found equal to $9.6 \times 10^{-5}\,(°\text{C})^{-1}$ (Figure (5)).

The measured pore pressure change with the temperature increase during the heating phase, presented in Figure (6), is corrected for the effect of the deformation of the drainage and pressure measurement systems, as explained in section 3.2.1, using equation (30) and the parameters presented in equation (31). The corrected thermal pressurization coefficient is evaluated as the slope of this curve and is equal to 0.62MPa/°C. The pressurization is stopped at the end of the heating phase, when the pore pressure approaches the confining pressure and the fluid leaks between the sample and the membrane. The comparison of the slopes of measured and corrected curves in Figure (6) shows that the corrected thermal pressurization coefficient is greater than the measured one. We also observe that this correction is more significant for the hardened cement paste than for example the Rothbach sandstone (Ghabezloo and Sulem [2]). The corrected curves are cut when the corrected pore pressure equals the confining pressure, at 54°C.

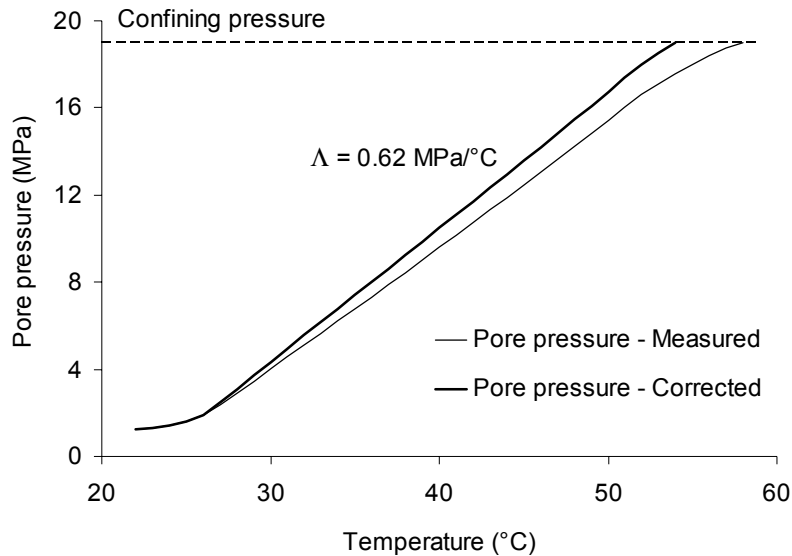

**Figure 6- Heating phase of the undrained heating test, measured and corrected pore pressure**

### 3.4.2 Cooling phase

The measured volumetric strain during cooling is presented on Figure (5). The measured thermal expansion at the beginning of the cooling phase is influenced by the presence of fluid between the sample and the membrane, as explained above. The undrained thermal expansion coefficient, which is the slope of the temperature-volumetric strain curve, then increases and remains constant during the test, equal to $1.2 \times 10^{-4} \left( °C \right)^{-1}$ which is slightly greater than the measured value during the heating phase.

Figure (7) presents the measured pore pressure during the cooling phase with the temperature. The corrected pore pressure curve is also presented on the same figure along with the calculated thermal pressurization coefficient. Due to the fluid leakage between the sample and the membrane at the end of the heating phase, the pore pressure decrease at the beginning of the cooling phase is not visible. The correction of the effect of the deformation of the drainage and pressure measurement systems resulted in a thermal pressurization coefficient which is here again greater than the measured one. The corrected thermal pressurization coefficient measured in the cooling phase is equal to 0.57MPa/°C.

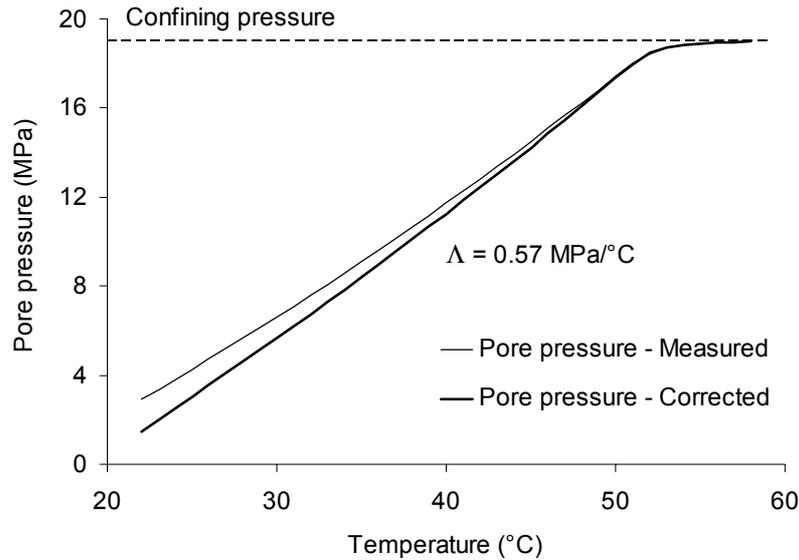

**Figure 7- Cooling phase of the undrained heating test, measured and corrected pore pressure**

# 4 Discussion of the results

In the tests results presented above, it is found that the thermal pressurization curve during heating and cooling is almost a straight line so that the thermal pressurization coefficient of the hardened cement paste can be considered as a constant, equal to 0.6MPa/°C. It has been checked that this value is retrieved when the test is repeated.

From equation (17) we can see that $\Lambda$ depends on the physical properties of water and on the thermal and mechanical properties of the porous material considered. Stress and temperature dependency of these parameters can thus lead to a variation of the thermal pressurization coefficient with the temperature and the level of stress. The variations of physical properties of pure water, $\alpha_f$ and $K_f$ with temperature and pressure are shown in Figure (8) [31]. We can see that $\alpha_f$ varies significantly with temperature but is less sensitive to pressure variations. On the other hand, in the range of temperature and pressure of our tests, $K_f$ does not vary much and takes values between 2.2GPa and 2.4GPa. The experimental study of Ghabezloo *et al*. [12] has shown the stress dependent character of the mechanical properties of the hardened cement paste. In this previous study, we have obtained that the elastic bulk modulus is reduced when the effective stress is increased as the result of microcracking. On the other hand, the degradation of the elastic modulus of the hardened cement paste with temperature increase is shown in the experimental studies of Dias *et al*. [32], Masse *et al*. [33] and Farage *et al*. [34]. DeJong [35] has also shown the degradation of the elastic properties of the hardened cement paste by performing nanoindentation tests at different temperatures. Then one can ask the question why is it found that the thermal pressurization coefficient is constant during the undrained heating test?

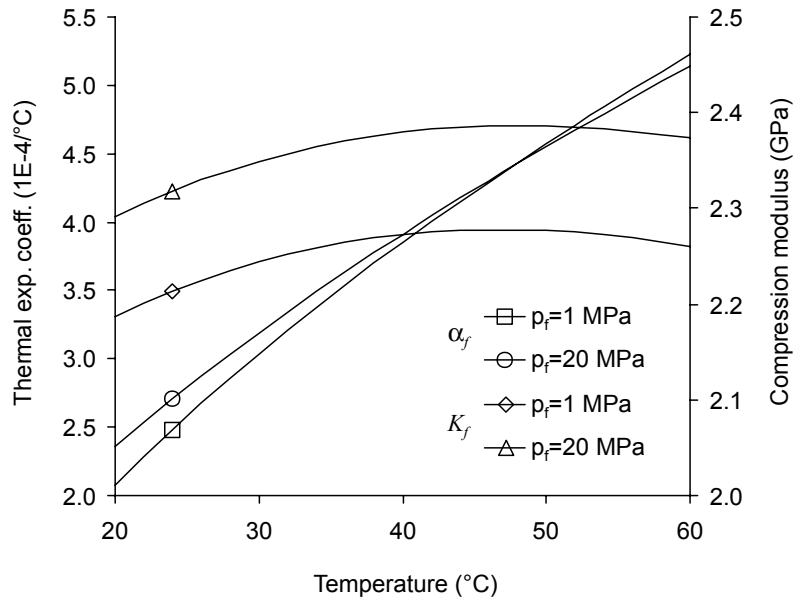

**Figure 8- Variations of the thermal expansion coefficient and of the compression modulus of pure water with temperature and pressure**

One possible explanation can be found in the anomalies of the thermal expansion of cement paste pore fluid, as presented by Valenza and Scherer [36]. They evaluated the permeability of the hardened cement paste using two different methods: thermopermeametry and beam bending. According to these authors, the comparison of the measurements using these methods showed that to bring the two measurements into agreement, the pore fluid in the fine pores of the hardened cement paste should have a thermal expansion coefficient about one and a half times larger than the one of the bulk liquid. Indeed, thermal expansion of water, when confined in nanopores is higher than that of bulk water. This phenomenon is showed experimentally by Derjaguin *et al.* [37] who studied the thermal expansion of water in nanopores of silica gel (5nm) and observed that the thermal expansion of water in small pores is anomalously higher than that of bulk water. The results presented by these authors also show that the rate of increase of the thermal expansion of water confined in nanopores with temperature is smaller than the one of bulk water. The ratio of the thermal expansion of water in nanopores to that of bulk water decreases with temperature. For temperatures higher than 70°C no more difference is observed between the thermal expansion coefficients. Xu *et al.* [38] also studied the thermal expansion and viscosity of water and salt solutions in porous silica glasses with two different pore sizes and found that the value of the thermal expansion of confined water is greater than that of bulk water. Additionally, the thermal expansion of water in smaller pores (5.0nm) is found to be higher than that in larger pores (7.4nm).

The possible effect of the anomalies of thermal expansion of water on our test results can be investigated by analysing the performed undrained heating test. In this test the confining pressure remains constant, $d\sigma = 0$, so that $d\sigma_d = -dp_f$. Inserting this expression in equation (9), the following expression is obtained:

$$dp_f = \frac{-1}{(1/K_d - 1/K_s)}(d\varepsilon + \alpha_d dT) \tag{32}$$

Inserting the term $1/K_d - 1/K_s$ from equation (32) in equation (15), with $d\sigma = 0$, we get

$$\alpha_f dT = \alpha_\phi dT - \frac{1}{\phi}(d\varepsilon + \alpha_d dT) + dp_f\left(\frac{1}{K_f} - \frac{1}{K_\phi}\right) \tag{33}$$

Replacing $d\varepsilon = -\alpha_u dT$ and $dp_f = \Lambda dT$ in equation (33) the following expression is obtained for the thermal expansion coefficient of the pore fluid:

$$\alpha_f = \alpha_\phi + \frac{1}{\phi}(\alpha_u - \alpha_d) + \Lambda\left(\frac{1}{K_f} - \frac{1}{K_\phi}\right) \tag{34}$$

Based on the results of Ghabezloo *et al.* [12], we take $K_\phi = 16.9\,\text{GPa}$ and we assume that this value does not vary during the test. Furthermore, in the absence of any information on the value of $\alpha_\phi$, we assume that $\alpha_\phi = \alpha_d = 6 \times 10^{-5}\,(°C)^{-1}$. In the analysis we take $\phi = 0.26$, $\Lambda = 0.6$ and $\alpha_u = 1.08 \times 10^{-4}\,(°C)^{-1}$ which are average values obtained from the heating and the cooling phase (Figures 5-7). The compression modulus $K_f$ of the pore fluid is taken equal to the one of pure water which as seen before does not vary significantly for the range of pressures and temperatures of the test $K_f = 2.27\,\text{GPa}$. With this set of values, from equation (34) we obtain $\alpha_f = 4.7 \times 10^{-4}(°C)^{-1}$. On the other hand, we can also evaluate $\alpha_f$ by using equation (20). The value of the Skempton coefficient $B$ is taken from the undrained compression tests of Ghabezloo *et al.* [12] at ambient temperature: $B = 0.4$. We obtain $\alpha_f = 5.2 \times 10^{-4}(°C)^{-1}$ which is in accordance with the one obtained above. These calculations give a constant value for the thermal expansion coefficient of the cement pore fluid which is obtained as an average value of the heating and cooling phases. A more detailed calculation can be done using equation (33) and the data of the performed undrained heating test. By doing so, the value of the thermal expansion coefficient of the cement pore fluid can be calculated for each data point during the test. The analysis is applied here for the cooling phase where the creep effects are of less importance. Moreover, the first part of the cooling phase in which the pore pressure reduction is delayed due to the presence of water between the sample and the membrane, is excluded from the analysis and only the temperatures below 45°C are considered. The result is shown in Figure (9). On the same graph the evolution of the thermal expansion coefficients of pure bulk water and of 0.5 mol/l NaOH bulk solution (data from Ref. [39]) are also plotted. On Figure (9), we observe that the thermal expansion of cement pore fluid is larger than the one of pure bulk water which is compatible with the results of Valenza and Scherer [36]. We observe also that the rate of increase of the thermal expansion of cement pore fluid with temperature is lower. This is compatible with the experimental results of Derjaguin *et al*. [37] and Xu *et al*. [38] who found that the rate of increase of the thermal expansion of water confined in nanopores of silica with temperature is smaller than the one of bulk water, as explained above. On the other hand this lower rate of increase with temperature is probably also due to the presence of dissolved ions in cement pore fluid. It is well-known that the presence of ions in water can influence its thermal expansion. Comparing the thermal expansion of pure bulk water with the one of 0.5mol/l NaOH bulk solution, we can see that the presence of ions increases the thermal expansion coefficient and decreases its rate of change with temperature. Additional results with various concentrations can be found in Valenza and Scherer [36] (their Figure 8). According to Taylor [23], typical concentrations after 180 days for pastes of w/c ratio 0.5 for $Na^+$, $K^+$ and $OH^-$ are respectively 0.08mol/l, 0.24mol/l and 0.32mol/l for low-alkali cement. The values for high-alkali cement are respectively equal to 0.16mol/l, 0.55mol/l and 0.71mol/l. Considering these concentrations, we use the thermal expansion of 0.5mol/l NaOH solution presented in Figure (9) as the thermal expansion of the bulk fluid in the cement paste. By doing so, we can evaluate the effect of the anomaly of pore fluid thermal expansion as ratio between the thermal expansion of the (confined) pore fluid and the one of the bulk fluid. This ratio reflects the effect of cement pore structure on the thermal expansion of pore fluid and is almost independent of presence of dissolved ions in the case of cement pore fluid which is mainly composed of univalent ions. According to experimental results of Xu *et al*. [38], the presence of dissolved ions in water can increase the anomaly of thermal expansion, but this influence is very small in the case of univalent ions. These authors measured the thermal expansion of the pure water and a 1.0mol/l NaCl solution for bulk fluids and when confined in silica pores (7.4nm). From their results we can evaluate the confined/bulk ratio of the thermal expansion of the NaCl solution which is only 2% larger than that of pure water. This result shows that the confined/bulk ratio of the thermal expansion of cement paste pore fluid is almost independent of the presence of dissolved ions and can be compared with that of pure water confined in silica pores. The comparison is presented in Figure (10) and shows a good accordance with the experimental results of Xu *et al*. [38]. This good accordance, when the effect of the dissolved ions is excluded from the results by using the confined/bulk thermal expansion ratio, clearly shows the effect of cement pore structure on the anomaly of cement paste pore fluid.

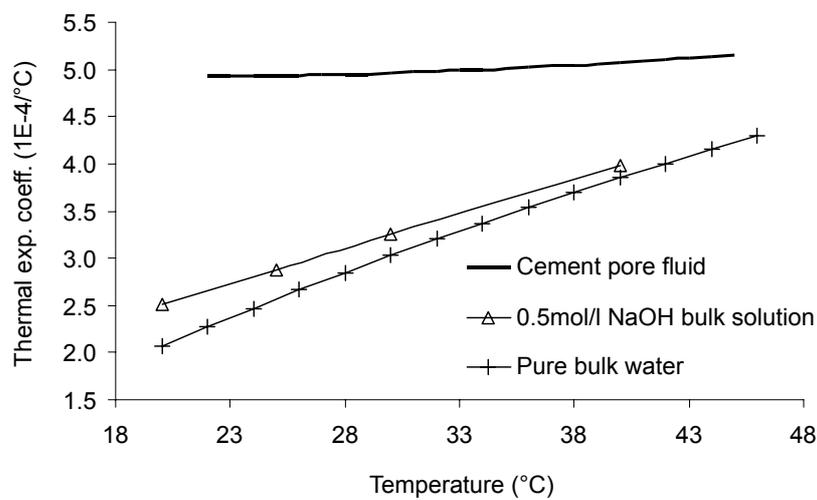

**Figure 9- Evaluated thermal expansion coefficient of cement pore fluid compared with the thermal expansion coefficients of pure water, and of 0.5 mol/l NaOH solution (data from [40])**

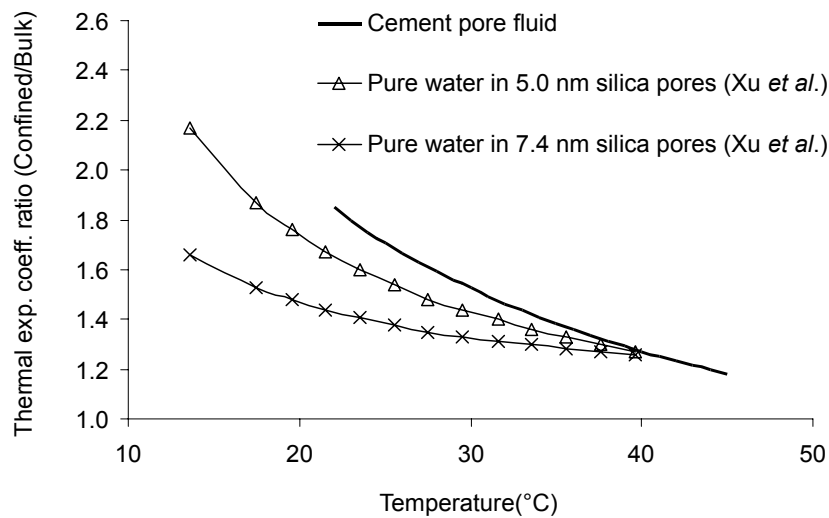

**Figure 10- Anomaly of the thermal expansion of cement pore fluid compared with pure water confined in silica pores of different size (data from Xu et al. [39])**

The above analysis shows several competing effects: (a) the increase of the thermal expansion coefficient with temperature, (b) the anomalous thermal behaviour of the confined pore fluid, (c) the decrease of this effect with temperature increase, (d) the effect of dissolved ions in the cement pore fluid. The combination of these effects may be a possible explanation for obtaining a thermal expansion coefficient of the pore fluid which does not vary significantly with temperature and consequently a constant thermal pressurization coefficient, as measured in the performed undrained heating test. We should also emphasized that the present analysis is based also on a set of assumptions: the variations of the poromechanical parameters of the hardened cement paste with the temperature and stress state are ignored, the thermal expansion coefficients $\alpha_\phi$ and $\alpha_d$ are assumed to be equal and the compression modulus $K_f$ of cement pore fluid is assumed to be equal to the one of pure bulk water. We would like also to point out that the low temperature dependency of cement paste pore fluid obtained from the back analysis of the results of the undrained heating test is not in agreement with some measurements of Xu *et al.*, presented by Valenza and Scherer [36] (Figure 9 in [36] ). Although the thermal expansion coefficient of the cement fluid presented by these authors is of the same range of our results, they obtain a stronger temperature dependency similar to that of pure bulk water. We believe that the experimental evaluation of the thermal properties of the cement paste fluid is still an open question which requires future investigation as it can only be addressed through indirect methods.

# 5 Conclusions

A better understanding of the effect of rapid temperature changes is an important point to properly evaluate the integrity of an oil well cement sheath submitted to such a variation. With this objective in mind, in this paper the effect of undrained heating on induced thermal pressurization and thermal expansion is formulated theoretically within the framework of thermo-poro-mechanics and is studied experimentally for a fluid-saturated hardened cement paste. The experimental study is performed on a hardened class G oilwell cement paste prepared with $w/c = 0.44$.

In the analysis of test results, special care is taken to the effect of the deformation of the drainage and pore pressure measurement systems and appropriate corrections are brought to the raw data. It is shown that these corrections can be significant and that the errors induced by the compressibility and the thermal deformation of the drainage systems should not be

Ghabezloo et al. (2008): The effect of undrained heating on a fluid-saturated hardened cement paste neglected. The drained thermal expansion coefficient of the hardened cement paste is measured in a drained heating test and found equal to $6 \times 10^{-5} \, (°C)^{-1}$. An undrained heating test with a heating-cooling cycle is performed and shows a proportional change of pore fluid pressure with temperature change. This linear response is described by a constant thermal pressurization coefficient equal to 0.6MPa/°C. It is expected that this coefficient varies during the heating and cooling phases due to the variations of the thermal expansion of water with temperature. The analysis of the results of the undrained heating test showed that this phenomenon may be attributed to the anomalies of the thermal expansion of cement paste pore fluid. The results of the undrained heating test are thus used to back analyse the thermal behaviour of cement pore fluid on the basis of the observation of a constant thermal pressurization coefficient in the range of temperature of the test. The evaluated thermal expansion of pore fluid is larger than the one of pure bulk water and its rate of increase with temperature is smaller. These anomalies are mainly attributed to the thermal behaviour of water when confined in small pores and also to the presence of dissolved ions in the cement paste pore fluid.

# 6   Appendix 1: Heating rate for drained and undrained heating tests

When performing elementary tests in the laboratory, the response of the sample should reproduce the response of a representative volume of the considered continuum. Thus, constitutive tests must be performed in such a way that temperature, strain and stress inside the sample can be assumed homogeneous. The choice of the loading rate is crucial to insure the homogeneity condition and preclude all kinetic effects.

For the undrained heating test, the rate of temperature change should be slow enough to ensure the homogeneity of temperature in the sample. Consequently, the induced pore pressure change inside the sample is quasi instantaneously homogeneous. For the drained heating test, in addition to the temperature homogeneity, the rate of temperature change should be slow enough to allow for the dissipation of the generated pore pressure change.

The rate of temperature change necessary for homogeneity of temperature in the sample can be analysed using the solution of temperature diffusion in an infinite cylinder of radius *R*. The initial condition at $t = 0^-$ is $T = T_0$ for $r \leq R$, where $t$ is the time and $r$ is the radial coordinate. At $t = 0^+$ the boundary condition is $T = T_1$ at $r = R$. Considering the dimensionless variable $\theta = (T - T_1)/(T_0 - T_1)$, the diffusion of temperature in the cylinder is given by the following differential equation:

$$\frac{\partial \theta}{\partial t} = C_T \left( \frac{\partial^2 \theta}{\partial r^2} + \frac{1}{r} \frac{\partial \theta}{\partial r} \right) \tag{35}$$

where $C_T$ is the thermal diffusivity. The solution of this differential equation can be expressed in terms of a series of Bessel functions $r$ and exponential functions for $t$:

$$\theta(r,t) = 2 \sum_{n=1}^{\infty} e^{-C_T \lambda_n^2 t} \frac{J_0(\lambda_n r)}{\lambda_n J_1(\lambda_n)} \tag{36}$$

where $J_0$ and $J_1$ are respectively Bessel functions of the first kind of order 0 and 1, and $\lambda_n$ are the roots of $J_0$. This solution is presented and used in a slightly different form also by Ciardullo *et al.* [40]. Taking $R = 19\,\text{mm}$, $T_0 = 20°C$ and $C_T = 0.33\,\text{mm}^2/\text{s}$ from Ref. [40], we consider an instantaneous temperature increase of 0.1°C at the exterior of the cylinder, $T_1 = 20.1°C$. Using equation (36) one finds that after only one second, the temperature at the centre of the sample is equal to 20.096°C. So one can see that the homogeneity of temperature in the sample is quasi instantaneous, and consequently the homogeneity of the pore pressure in the undrained heating test is also quasi instantaneous. Therefore, in order to have a good homogeneity of the temperature between the heating element at the exterior part of the triaxial cell and the sample's temperature, a heating rate of 0.1°C/min is chosen for the undrained heating test.

Now assuming instantaneous temperature homogeneity in the sample, the rate of temperature change for the drained heating test can be analysed using the problem of one dimensional pressure diffusion along the axis of the sample. In order to derive this equation, the fluid mass conservation in a deforming elementary volume reads:

$$\frac{dm_f}{dt} + m_f \frac{dV/V}{dt} + \frac{\partial q}{\partial z} = 0 \tag{37}$$

where $m_f$ is the fluid mass per unit volume of the porous material, $m_f = \rho_f \phi$ and $q$ is the fluid mass flux per surface area (orthogonal to flow direction $z$). Knowing that $dm_f = \phi d\rho_f + \rho_f d\phi$, $d\rho_f = \rho_f \left( dp_f/K_f - \alpha_f dT \right)$ and equation (10) one obtains:

*Ghabezloo et al. (2008): The effect of undrained heating on a fluid-saturated hardened cement paste*

$$dm_f = \phi \rho_f \left( \frac{dp_f}{K_f} - \alpha_f dT + \frac{dV_\phi}{V_\phi} - \frac{dV}{V} \right) \tag{38}$$

Inserting equation (38) and $m_f = \rho_f \phi$ in equation (37) the following expression is obtained:

$$\phi \rho_f \left( \frac{1}{K_f} \frac{dp_f}{dt} - \alpha_f \frac{dT}{dt} + \frac{dV_\phi/V_\phi}{dt} \right) + \frac{\partial q}{\partial z} = 0 \tag{39}$$

Replacing equation (8) and then equation (7) in equation (39) the following relation is found:

$$\left[ \phi \left( \frac{1}{K_f} - \frac{1}{K_\phi} \right) + \left( \frac{1}{K_d} - \frac{1}{K_s} \right) \right] \frac{dp_f}{dt} - \left( \frac{1}{K_d} - \frac{1}{K_s} \right) \frac{d\sigma}{dt} - (\alpha_f - \alpha_\phi) \frac{dT}{dt} + \frac{1}{\rho_f} \frac{\partial q}{\partial z} = 0 \tag{40}$$

Using equations (16) and (17), equation (40) can be written in the following form:

$$\frac{dp_f}{dt} = B \frac{d\sigma}{dt} + \Lambda \frac{dT}{dt} - \frac{\beta_u}{\rho_f} \frac{\partial q}{\partial z} = 0 \tag{41}$$

where

$$\beta_u = \frac{1}{\phi(1/K_f - 1/K_\phi) + (1/K_d - 1/K_s)} \tag{42}$$

The fluid mass flux $q$ in equation (41) is given by Darcy's law:

$$q = -k \frac{\rho_f}{\mu_f} \frac{\partial p_f}{\partial z} \tag{43}$$

where $k$ is the permeability and $\mu_f$ is the fluid viscosity. Replacing equation (43) in equation (41) the following expression is obtained:

$$\frac{dp_f}{dt} = B \frac{d\sigma}{dt} + \Lambda \frac{dT}{dt} + \frac{\beta_u}{\rho_f} \frac{\partial}{\partial z} \left( k \frac{\rho_f}{\mu_f} \frac{\partial p_f}{\partial z} \right) = 0 \tag{44}$$

Equation (44) can be extended to take into account the effect of non-elastic strains as presented in section 2.1. The effect of creep is also presented in Ref. [41]. By doing so, in the case of a material for which the solid phase is elastic ($d\varepsilon_s^{ne} = 0$), one obtains:

$$\frac{dp_f}{dt} - \beta_u \frac{d\varepsilon^{ne}}{dt} = B \frac{d\sigma}{dt} + \Lambda \frac{dT}{dt} + \frac{\beta_u}{\rho_f} \frac{\partial}{\partial z} \left( k \frac{\rho_f}{\mu_f} \frac{\partial p_f}{\partial z} \right) = 0 \tag{45}$$

The drained heating test is performed under constant confining pressure, so that $d\sigma/dt = 0$. In order to exclude the effect of creep deformations during the test, the test is performed under a low confining pressure, so that $d\varepsilon^{ne}/dt = 0$. From equation (45) one can see that the diffusion of pressure in the sample depends directly on the permeability $k$, the thermal pressurization coefficient $\Lambda$ of the tested sample and the length of the drainage path. The permeability of the tested sample is evaluated equal to $1.1 \times 10^{-19}$ m² [42]. The elastic moduli required for evaluation of $\beta_u$ in equation (42) are evaluated experimentally by Ghabezloo *et al.* [12] at ambient temperature: $K_d = 8.7\text{GPa}$, $K_s = 21\text{GPa}$, $K_\phi = 16.9\text{GPa}$, $\phi = 0.26$. Based on some preliminary calculations, the heating rate of 0.08°C/min is chosen for the drained heating test. Compared to the rates of temperature change used by Helmuth [43] and Sabri and Illston [44], 0.25°C/min and 0.33°C/min respectively, the chosen rate of temperature change, 0.08°C/min, can be considered to be relatively slow. In the following, we show that this choice of the heating rate is satisfactory. In this analysis, we use the value of thermal pressurization coefficient $\Lambda$, equal to 0.6MPa/°C, which is obtained in the performed undrained heating test, as presented in section 3.4. A numerical simulation of drained heating test is performed in which the temperature of the sample is increased with different rates from 20°C to 50°C. Only one half of the tested sample is modelled which makes the model length equal to 38mm. At one end of the sample, a constant pore pressure equal to 1.0MPa is applied, while at the other end a no-flow condition is imposed. Equation (45) is solved using a finite difference scheme with the parameters presented above. The physical properties of water, density, compressibility and viscosity, are assumed equal to the ones of pure water and their variations with temperature are taken into account. Figure (11) presents the distribution of the pore pressure along the height of the sample at the end of the numerical heating test, for three different heating rates. One can see that for the heating rate used in this study, 0.08°C/min, a relatively small pressure difference equal to 0.43MPa remains along the axis of the sample. Using equation (9), $K_d = 8.7\text{GPa}$ and $K_s = 21\text{GPa}$, this pore pressure difference of 0.43MPa induces a difference of 30μm/m (0.003%) in the volumetric strains between the middle and the end of the sample. Obviously this small strain heterogeneity has a negligible effect in the evaluation of the thermal expansion coefficient. Notice that with $\Lambda = 0.6\,\text{MPa}/°\text{C}$, an undrained heating test from 20°C to 50°C would result in an excess pore pressure of 18MPa. This excess pore pressure in undrained conditions can be compared with the excess pore pressure obtained in the performed numerical drained heating test, equal to 0.43MPa, which shows that a heating rate, 0.08°C/min as used in our experiments guaranties that the heating test is performed in drained condition. It is interesting to note that, due to the relatively low permeability of the sample and the high value of $\Lambda$, even by using a r heating rate four times slower, 0.02°C/min, a pressure difference of 0.17MPa still remains inside the sample. This pressure difference increases to 1.14MPa for a heating rate of 0.20°C/min (Figure 11).

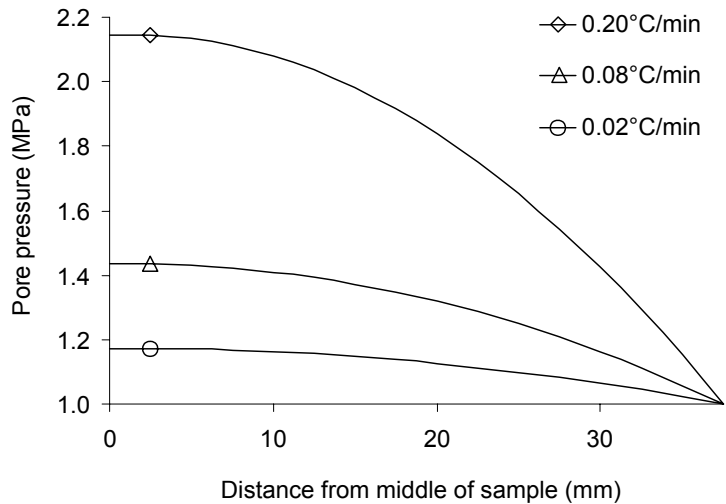

**Figure 11- Numerical simulation of a drained heating test from 20°C to 50°C with different heating rates: Pore pressure along the height of the sample**

# 7    Acknowledgments

The authors gratefully acknowledge TOTAL for supporting this research. They wish also to thank George W. Scherer for interesting discussions and François Martineau for his assistance in the experimental work.